\documentclass[global]{svjour}
\usepackage{amsfonts}
\usepackage{amsmath}

\input{tcilatex}

\begin{document}

\journalname{Journal of Mathematical Physics}
\title{Solution of the Dirac equation in the rotating Bertotti-Robinson
spacetime}
\author{A. Al-Badawi \inst{1} \and I. Sakalli\inst{2}}
\institute{Department of Physics, AL-Hussein Bin Talal University, Ma'an, Jordan 
\email{dr.ahmadbadawi@ahu.edu.jo}%
\and Department of Physics, Eastern Mediterranean University, Gazimagosa,
North Cyprus, Mersin 10, Turkey 
\email{izzet.sakalli@emu.edu.tr}%
.}
\dedication{}
\offprints{}
\mail{}
\maketitle

\begin{abstract}
The Dirac equation is solved in the rotating Bertotti-Robinson spacetime.
The set of equations representing the Dirac equation in the Newman-Penrose
formalism is decoupled into an axial and angular part. The axial equation,
which is independent of mass, is solved exactly in terms of hypergeometric
functions. The angular equation is considered both for massless (neutrino)
and massive spin-$\frac{1}{2}$ particles. For the neutrinos, it is shown
that the angular equation admits an exact solution in terms of the confluent
Heun equation. In the existence of mass, the angular equation does not allow
an analytical solution, however, it is expressible as a set of first order
differential equations apt for numerical study.
\end{abstract}


\section{Introduction}

The Rotating Bertotti-Robinson spacetime (RBR), which was discovered long
time ago by Carter \cite{R1}, is an Einstein-Maxwell solution representing a
rotating electromagnetic field. This solution remained unnoticed until the
study of Al-Badawi and Halilsoy \cite{R2}, who rediscovered it by applying a
coordinate transformation to the cross-polarized Bell-Szekeres solution of
colliding electromagnetic waves \cite{R3}. We can consider the RBR solution
as an extended version of the Bertotti-Robinson (BR) solution due to the
fact that the RBR solution contains one more degree of freedom to be
interpreted as rotation. Adding rotation to the BR creates gravitational
curvature, distorts isotropy and modifies geodesics significantly. For this
reason the RBR solution assumes more complicated topology compared with the
BR solution. The RBR spacetime has the topology of $AdS^{2}\times S^{2}$ and
underlying group structure of $SL(2,R)\times U(1)$. Nowadays, the spacetimes
with AdS structure are quite popular because of their connection with string
theory, higher dimensions and the brane worlds. The RBR solution can also be
interpreted as the "throat" connecting two rotating black holes with
charges. This is due to the fact that the BR solution is considered as the
"throat" connecting two asymptotically flat Reissner-Nordstr\"{o}m regions 
\cite{R4}.

The first study on the Dirac equation in the BR spacetime without charge
coupling was considered long time ago \cite{R5}. During the last decade the
studies on spin-$\frac{1}{2}$ particles in the BR spacetime gained momentum.
For example, Silva-Ortigoza \cite{R6} showed how the Dirac equation could be
separated when the background is the BR spacetime with cosmological
constant. Later on, a more detailed study on the problem of the Dirac
equation in the BR spacetime was worked out as well \cite{R7}. Here, we
extend this recent work by looking for the answer to the following question
: "How does a test Dirac particle behave in a rotating spacetime filled with
electromagnetic field i.e., in the RBR spacetime?". We shall ignore the
back-reaction effect of the spin-$\frac{1}{2}$ particle on the spacetime, by
the same token as done in \cite{R7}. The RBR solution represents one of the
type-D spacetimes and as it could be followed from the literature the
studies on spin-$\frac{1}{2}$ particles in type-D spacetimes have always
attracted attention \cite{R8} (and references therein). More recently, the
problem of Dirac equation in the near horizon geometry of an extreme Kerr
black hole (Kerr throat) has been studied by \cite{R9}. In many aspects,
that spacetime of the Kerr throat shares common features with the RBR
spacetime. The main difference between them is that the Kerr throat is a
vacuum solution, while the RBR is not. On the other hand, they are both
regular solutions.

In this paper, our aim is to solve the Dirac equation in the RBR spacetime.
To this end in order to separate the Dirac equation we employ the well-known
method suggested by Chandrasekhar \cite{R8}. We separate the Dirac equation
into the axial (function of $z$ only) and the angular parts (function of $%
\theta $ only) in such a way that the resulting axial equation remains
independent of mass. This advantage leads us to obtain an exact solution for
the axial equations in terms of hypergeometric functions. The angular part
turns out to be more complicated than the axial part. This is due to the
fact that the metric functions are dependent on the variable $\theta ,$ and
also the angular equations contain the mass term. For the angular part two
separate cases, massive and massless (neutrinos) cases are discussed. The
angular equations of the massive case, we are able to reduce the equations
to a set of linear first order differential equations, which can be utilized
numerically. However the massless particle (neutrino) equations are solved
exactly by reducing the equations to the confluent Heun equations.

Recall that, the confluent Heun differential equations are less known than
the hypergeometric family in literature. The modern mathematical development
shows that many physical problems are solved exactly by Heun functions. \cite%
{R10,R11,R12}. For example, problems involving atomic physics with certain
potentials \cite{R13} which combine different inverse power or combining the
quadratic potential with inverse powers of two, four, and six etc.. Problems
in solid state physics, like dislocation movement in crystalline materials
and quantum diffusion of kinks along dislocations are also solved in terms
of Heun function \cite{R14}. Problems in general relativity, Fiziev \cite%
{R15} gave an exact solution of the Regge-Wheeler equation in terms of the
Heun functions and apply them for study of different boundary problems. More
recently, Birkandan and Hortacsu \cite{R16} have given examples in which the
Heun functions admit the solution of the wave equation encountered in
general relativity. They have related the solutions of the Dirac equation,
when the background is the Nutku's helicoid spacetime in five dimensions to
the double confluent Heun function. Nowadays, the modern computer packages
have started to involve the Heun functions in their algorithms, as for
instance it can be seen in the $10^{th}$ and higher versions of the famous
computer package, Maple.

The paper is organized as follows: In section II, a brief review of the RBR
solution is given. Next, we present the basic Dirac equations and separate
them in the spacetime of RBR. In section III, we present the exact solution
of the axial equation. The angular equation with both massless and massive
cases are discussed in sections IV and V, respectively. Finally, in section
VI, we draw our conclusions.

\bigskip

\section{Rotating Bertotti-Robinson spacetime and separation of Dirac
equation on it}

The metric describing a rotating electromagnetic field, RBR solution,
written in spherical coordinates, is given by \cite{R2} 
\begin{equation}
ds^{2}=\frac{F(\theta )}{r^{2}}\left[ d\widetilde{t}^{2}-dr^{2}-r^{2}d\theta
^{2}-\frac{r^{2}\sin ^{2}\theta }{F^{2}}\left( d\widetilde{\phi }-\frac{q}{r}%
d\widetilde{t}\right) ^{2}\right] ,  \label{1}
\end{equation}

where the function $F(\theta)$\ and the constant $q$\ are%
\begin{equation*}
F(\theta)=1+a^{2}\left( 1+\cos^{2}\theta\right) ,
\end{equation*}%
\begin{equation}
q=2a\sqrt{1+a^{2}}.  \label{2}
\end{equation}

in which $a$ is the rotation parameter. It is readily seen that for $a=0$
metric (1) reduces to the BR metric \cite{R7} (and references cited therein).

We make the choice of the following null tetrad basis 1-forms $\left( l,%
\text{ }n,\text{ }m,\text{ }\overline{m}\right) $ of the Newman-Penrose (NP)
formalism \cite{R17} in terms of the RBR geometry that satisfies the
orthogonality conditions, $l.n=-m.\overline{m}=1.$ We note that throughout
the paper, a bar over a quantity denotes complex conjugation. We can write
the covariant 1-forms as 
\begin{equation*}
\sqrt{2}l=\frac{1}{2r}\sqrt{F}\left( d\widetilde{t}-dr\right) ,
\end{equation*}%
\begin{equation*}
\sqrt{2}n=\frac{2}{r}\sqrt{F}\left( d\widetilde{t}+dr\right) ,
\end{equation*}%
\begin{equation}
\sqrt{2}m=i\sqrt{F}d\theta +\frac{\sin \theta }{\sqrt{F}}\left( \frac{2a}{r}%
\sqrt{1+a^{2}}d\widetilde{t}-d\widetilde{\phi }\right) .  \label{3}
\end{equation}

We obtain the non-zero $\Psi _{2}$ and $\Phi _{11}$, which are known as Weyl
and Maxwell scalars, respectively, as 
\begin{equation*}
\Psi _{2}=\frac{a^{2}}{F^{3}}\left[ \left( 1+a^{2}\right) \cos 2\theta
+a^{2}\cos ^{2}\theta -\frac{i}{a}\sqrt{1+a^{2}}(1+a^{2}+a^{2}\sin
^{2}\theta )\cos \theta \right] ,
\end{equation*}%
\begin{equation}
\Phi _{11}=\frac{1}{2F^{2}}.  \label{4}
\end{equation}%
\ \ \ 

The singularity free \ and the type-D characters of the metric can be easily
deduced from $\Psi _{2}$. It is obvious that for $a=0,$ this type-D metric
(1) yields a conformally flat spacetime (i.e. the BR spacetime) in which a
uniform electromagnetic field, with $\Phi _{11}=\frac{1}{2},$ fills the
entire space. As it can be seen from Eq. (4), rotation ($a\neq 0$) gives
rise to anisotropy of the prevailing electromagnetic field.

In order to study the Dirac equation in the RBR spacetime we prefer to work
in a more convenient coordinate system, therefore by using the following
transformation, 
\begin{equation*}
z=\frac{1}{2r}(\widetilde{t}^{2}-r^{2}+1),
\end{equation*}

\begin{equation*}
t=\tan^{-1}\left[ \frac{1}{2\widetilde{t}}(\widetilde{t}^{2}-r^{2}-1)\right]
,
\end{equation*}

\begin{equation}
\phi=\frac{1}{2}q\ln(\frac{(r-\widetilde{t})^{2}+1}{(r+\widetilde{t})^{2}+1}%
)+\widetilde{\phi},  \label{5}
\end{equation}

metric (1) is transformed into 
\begin{equation}
ds^{2}=F(\theta )\left[ (1+z^{2})dt^{2}-\frac{dz^{2}}{(1+z^{2})}-d\theta
^{2}-\frac{\sin ^{2}\theta }{F^{2}}\left( d\phi -qzdt\right) ^{2}\right] .
\label{6}
\end{equation}

We notice that the metric functions in Eq. (6) depend explicitly on the
variable $\theta $ as a result the angular part of the Dirac equation
becomes important. The coordinates $-\infty <t<\infty ,$ $-\infty <z<\infty $
covers the entire, singularity free spacetime.

The covariant 1-forms of the metric (6) can be taken as, 
\begin{equation*}
\sqrt{2}l=\sqrt{F}(\sqrt{1+z^{2}}dt-\frac{dz}{\sqrt{1+z^{2}}}),
\end{equation*}%
\begin{equation*}
\sqrt{2}n=\sqrt{F}(\sqrt{1+z^{2}}dt+\frac{dz}{\sqrt{1+z^{2}}}),
\end{equation*}%
\begin{equation}
\sqrt{2}m=\sqrt{F}d\theta+\frac{i\sin\theta}{\sqrt{2}\sqrt{F}}(d\phi-qzdt),
\label{7}
\end{equation}

while their corresponding directional derivatives become 
\begin{equation*}
\sqrt{2}D=\frac{\partial _{t}}{\sqrt{F}\sqrt{1+z^{2}}}+\frac{\sqrt{1+z^{2}}%
\partial _{z}}{\sqrt{F}}+\frac{qz\partial _{\phi }}{\sqrt{F}\sqrt{1+z^{2}}},
\end{equation*}%
\begin{equation*}
\sqrt{2}\Delta =\frac{\partial _{t}}{\sqrt{F}\sqrt{1+z^{2}}}-\frac{\sqrt{%
1+z^{2}}\partial _{z}}{\sqrt{F}}+\frac{qz\partial _{\phi }}{\sqrt{F}\sqrt{%
1+z^{2}}},
\end{equation*}%
\begin{equation*}
\sqrt{2}\delta =-\left[ \frac{\partial _{\theta }}{\sqrt{F}}+i\frac{\sqrt{F}%
}{\sin \theta }\partial _{\phi }\right] ,
\end{equation*}%
\begin{equation}
\sqrt{2}\bar{\delta}=-\left[ \frac{\partial _{\theta }}{\sqrt{F}}-i\frac{%
\sqrt{F}}{\sin \theta }\partial _{\phi }\right] ,  \label{8}
\end{equation}

Using the above tetrad we determine the nonzero NP complex spin coefficients 
\cite{R17} as, 
\begin{equation*}
\tau =-\pi =\frac{-1}{2\sqrt{2}F^{\frac{3}{2}}}\left[ a^{2}\sin (2\theta
)+iq\sin \theta )\right] ,
\end{equation*}

\begin{equation*}
\epsilon=\gamma=\frac{z}{2\sqrt{2}\sqrt{F}\sqrt{1+z^{2}}},
\end{equation*}

\begin{equation}
\alpha=-\beta=\frac{1}{4\sqrt{2}F^{\frac{3}{2}}}\left[ 2\cot\theta
(1+2a^{2})+iq\sin\theta\right] .  \label{9}
\end{equation}

The Dirac equations in the NP formalism are given by \cite{R8} as 
\begin{equation*}
\left( D+\epsilon-\rho\right) F_{1}+\left( \overline{\delta}+\pi
-\alpha\right) F_{2}=i\mu_{p}G_{1},
\end{equation*}%
\begin{equation*}
\left( \delta+\beta-\tau\right) F_{1}+\left( \Delta+\mu-\gamma\right)
F_{2}=i\mu_{p}G_{2},
\end{equation*}%
\begin{equation*}
\left( D+\overline{\epsilon}-\overline{\rho}\right) G_{2}-\left( \delta+%
\overline{\pi}-\overline{\alpha}\right) G_{1}=i\mu_{p}F_{2,}
\end{equation*}%
\begin{equation}
\left( \Delta+\overline{\mu}-\overline{\gamma}\right) G_{1}-\left( \overline{%
\delta}+\overline{\beta}-\overline{\overline{\tau}}\right)
G_{2}=i\mu_{p}F_{1},  \label{10}
\end{equation}

where $\mu^{\ast}=\sqrt{2}\mu_{p}$ is the mass of the Dirac particle.

The form of the Dirac equation suggests that we assume \cite{R8}, 
\begin{equation}
F_{1}=f_{1}\left( z\right) A_{1}\left( \theta\right) e^{i\left(
kt+m\phi\right) },  \notag
\end{equation}%
\begin{equation*}
F_{2}=f_{2}\left( z\right) A_{3}\left( \theta\right) e^{i\left(
kt+m\phi\right) },
\end{equation*}%
\begin{equation*}
G_{1}=g_{1}\left( z\right) A_{2}\left( \theta\right) e^{i\left(
kt+m\phi\right) },
\end{equation*}%
\begin{equation}
G_{2}=g_{2}\left( z\right) A_{4}\left( \theta\right) e^{i\left(
kt+m\phi\right) }.  \label{11}
\end{equation}

Here, we consider the corresponding Compton wave of the Dirac particle as in
the form of $f\left( z\right) A\left( \theta\right) e^{i\left(
kt+m\phi\right) },$ where $k$\ is the frequency of the incoming wave and $m$
is the azimuthal quantum number\ of the wave. The temporal and azimuthal
dependencies are chosen same but axial and angular dependencies are chosen
different for different spinors.

Substituting the appropriate spin coefficients (9) and the spinors (11) into
the Dirac equation (10), we obtain%
\begin{equation*}
\frac{\widetilde{Z}f_{1}}{f_{2}}-\frac{LA_{3}}{A_{1}}=i\mu^{\ast}\frac{g_{1}%
}{f_{2}}\frac{A_{2}}{A_{1}}\sqrt{F},
\end{equation*}%
\begin{equation*}
\frac{\overline{\widetilde{Z}}f_{2}}{f_{1}}+\frac{L^{\dag}A_{1}}{A_{3}}%
=-i\mu^{\ast}\frac{g_{2}}{f_{1}}\frac{A_{4}}{A_{3}}\sqrt{F},
\end{equation*}%
\begin{equation*}
\frac{\widetilde{Z}g_{2}}{g_{1}}+\frac{\pounds ^{\dag}A_{2}}{A_{4}}=i\mu
^{\ast}\frac{f_{2}}{g_{1}}\frac{A_{3}}{A_{4}}\sqrt{F},
\end{equation*}%
\begin{equation}
\frac{\overline{\widetilde{Z}}g_{1}}{g_{2}}-\frac{\pounds A_{4}}{A_{2}}%
=-i\mu^{\ast}\frac{f_{1}}{g_{2}}\frac{A_{1}}{A_{2}}\sqrt{F},  \label{12}
\end{equation}

where the axial and the angular operators, respectively, are

\begin{equation*}
\widetilde{Z}=\sqrt{1+z^{2}}\partial_{z}+\frac{1}{2\sqrt{1+z^{2}}}\left[
z+2i(k+mqz)\right] ,
\end{equation*}

\begin{equation*}
\overline{\widetilde{Z}}=\sqrt{1+z^{2}}\partial_{z}+\frac{1}{2\sqrt{1+z^{2}}}%
\left[ z-2i(k+mqz)\right] ,
\end{equation*}

and

\begin{equation*}
L=\partial_{\theta}+\frac{\cot\theta}{2}-\frac{a^{2}\sin2\theta}{4F}+\frac {%
mF}{\sin\theta}-i\frac{q\sin\theta}{4F},
\end{equation*}

\begin{equation*}
L^{\dag}=\partial_{\theta}+\frac{\cot\theta}{2}-\frac{a^{2}\sin2\theta}{4F}-%
\frac{mF}{\sin\theta}-i\frac{q\sin\theta}{4F},
\end{equation*}

\begin{equation*}
\pounds =\partial_{\theta}+\frac{\cot\theta}{2}-\frac{a^{2}\sin2\theta}{4F}+%
\frac{mF}{\sin\theta}+i\frac{q\sin\theta}{4F},
\end{equation*}

\begin{equation}
\pounds ^{\dag}=\partial_{\theta}+\frac{\cot\theta}{2}-\frac{%
a^{2}\sin2\theta }{4F}-\frac{mF}{\sin\theta}+i\frac{q\sin\theta}{4F}.
\label{13}
\end{equation}

It is obvious that \ $L$ and $L^{\dag}$ are purely angular operators and $L=%
\overline{\pounds },$ $L^{\dag}=\overline{\pounds }^{\dag}.$

In order to separate the Dirac equation (12) into axial and angular parts we
choose $f_{1}=g_{2}$, $f_{2}=g_{1},$ $A_{2}=\overline{A_{1}}$ and $A_{4}=%
\overline{A_{3}}$ and introduce a real separation constant $\lambda $ as 
\begin{equation}
\widetilde{Z}g_{2}=-\lambda g_{1},  \label{14}
\end{equation}%
\begin{equation}
\overline{\widetilde{Z}}g_{1}=-\lambda g_{2}.  \label{15}
\end{equation}

and

\begin{equation}
LA_{3}+i\mu^{\ast}\overline{A}_{1}\sqrt{F}=-\lambda A_{1},  \label{16}
\end{equation}

\begin{equation}
L^{\dag}A_{1}+i\mu^{\ast}\overline{A}_{3}\sqrt{F}=\lambda A_{3}.  \label{17}
\end{equation}

\section{Solution of the axial equation}

The structure of the axial Eqs. (14, 15) admits $g_{1}=\overline{g}_{2}$.
Thus it is enough to decouple the axial equations for $g_{2},$ namely

\begin{equation}
\overline{\widetilde{Z}}\left( \widetilde{Z}g_{2}\right) =\lambda^{2}g_{2},
\label{18}
\end{equation}

The explicit form of Eq. (18) can be obtained as

\begin{equation*}
(1+z^{2})g_{2}^{\prime \prime }\left( z\right) +2zg_{2}^{\prime }\left(
z\right) +\frac{1}{2(1+z^{2})}\left[ 1+\frac{1}{2}z^{2}+\right.
\end{equation*}

\begin{equation}
\left. 2(k+mqz)^{2}-2i(kz-mq)-2\lambda ^{2}(1+z^{2})\right] g_{2}\left(
z\right) =0,  \label{19}
\end{equation}

(Throughout the paper, a prime denotes the derivative with respect to its
argument.)

Let us introduce a new variable $y$ such that $z=i(1-2y)$, therefore Eq.
(19) becomes,%
\begin{equation*}
y(1-y)g_{2}^{\prime \prime }\left( y\right) +(1-2y)g_{2}^{\prime }\left(
y\right) -
\end{equation*}

\begin{equation*}
\frac{1}{4y(1-y)}\left\{ \frac{1}{4}-m^{2}q^{2}(1-2y)^{2}+y(1-y)+\right.
\end{equation*}

\begin{equation}
\left. k(k+1-2y)-4\lambda^{2}y(1-y)+imq\left[ 1+2k(1-2y)\right] \right\}
g_{2}\left( y\right) =0,  \label{20}
\end{equation}

\bigskip The exact solution of the axial part is found in terms of the Gauss
hypergeometric functions as,

\begin{align}
g_{2}(y)& =C_{1}y^{\alpha }(y-1)^{\beta }\text{ }_{2}F_{1}(\frac{1}{2}%
+k-\gamma ,\frac{1}{2}+k+\gamma ,\frac{3}{2}+k+imq;y)+  \notag \\
& C_{2}y^{-\alpha }(y-1)^{\beta }\text{ }_{2}F_{1}(-(\gamma +imq),\gamma
-imq,\frac{1}{2}-k-imq;y),  \label{21}
\end{align}

where the parameters are

\begin{equation*}
\alpha=\frac{1}{2}(k+\frac{1}{2}+imq),
\end{equation*}

\begin{equation*}
\beta=\frac{1}{2}(k-\frac{1}{2}-imq),
\end{equation*}

\begin{equation}
\gamma=\sqrt{\lambda^{2}-m^{2}q^{2}}.  \label{22}
\end{equation}

and $C_{1},C_{2}$ are complex constants.

\section{Reduction of the angular equation to Heun equation: The massless
case}

The aim of this section is to show that the angular Eqs. (16, 17) for the
massless Dirac particles (such as neutrinos) can be decoupled to a confluent
Heun differential equation.

For $\mu ^{\ast }=0,$ Eqs. (16, 17) can be explicitly\ written as

\begin{equation}
A_{1}^{\prime}\left( \theta\right) +\left( K-M\right) A_{1}\left(
\theta\right) =\lambda A_{3}\left( \theta\right) ,  \label{23}
\end{equation}%
\begin{equation}
A_{3}^{\prime}\left( \theta\right) +\left( K+M\right) A_{3}\left(
\theta\right) =-\lambda A_{1}\left( \theta\right) ,  \label{24}
\end{equation}

where

\begin{equation}
K=\frac{\cot\theta}{2}-\frac{a^{2}\sin2\theta}{4F}-i\frac{q\sin\theta}{4F},
\label{25}
\end{equation}

\begin{equation}
M=\frac{mF}{\sin\theta},  \label{26}
\end{equation}

Introducing the scalings 
\begin{equation}
A_{1}\left( \theta\right) =H_{1}\left( \theta\right) e^{-\int\left(
K-M\right) d\theta},  \label{27}
\end{equation}%
\begin{equation}
A_{3}\left( \theta\right) =H_{3}\left( \theta\right) e^{-\int\left(
K+M\right) d\theta},  \label{28}
\end{equation}
we get%
\begin{equation}
H_{1}^{\prime}\left( \theta\right) =\lambda H_{3}\left( \theta\right)
e^{-2\int Md\theta},  \label{29}
\end{equation}%
\begin{equation}
H_{3}^{\prime}\left( \theta\right) =-\lambda H_{1}\left( \theta\right)
e^{2\int Md\theta},  \label{30}
\end{equation}

Decoupling Eqs. (29, 30) in Eq. (29) for $H_{1}\left( \theta \right) ,$ we
obtain 
\begin{equation}
H_{1}^{\prime \prime }\left( \theta \right) +2MH_{1}^{\prime }\left( \theta
\right) +\lambda ^{2}H_{1}\left( \theta \right) =0,  \label{31}
\end{equation}

Similarly, if we decouple the axial equations for $H_{3}\left( \theta
\right) $, the resulting equation turns out to be 
\begin{equation}
H_{3}^{\prime \prime }\left( \theta \right) -2MH_{3}^{\prime }\left( \theta
\right) +\lambda ^{2}H_{3}\left( \theta \right) =0,  \label{32}
\end{equation}

Introducing a new variable $\theta =\cos ^{-1}\left( 1-2z\right) ,$ Eqs.
(31, 32) are cast into the general confluent form of Heun equation, namely 
\begin{equation}
H_{1}^{\prime \prime }\left( z\right) +\left[ -4a^{2}m+\frac{\frac{1}{2}%
+m+2a^{2}m}{z}+\frac{\frac{1}{2}-(m+2a^{2}m)}{z-1}\right] H_{1}^{\prime
}\left( z\right) -\frac{\lambda ^{2}}{z\left( z-1\right) }H_{1}\left(
z\right) =0,  \label{33}
\end{equation}%
\begin{equation}
H_{3}^{\prime \prime }\left( z\right) +\left[ 4a^{2}m+\frac{\frac{1}{2}%
-(m+2a^{2}m)}{z}+\frac{\frac{1}{2}+m+2a^{2}m}{z-1}\right] H_{3}^{\prime
}\left( z\right) -\frac{\lambda ^{2}}{z\left( z-1\right) }H_{3}\left(
z\right) =0.  \label{34}
\end{equation}

Recall that the general confluent form of Heun equation \cite{R12} is given
as follows 
\begin{equation}
H^{\prime \prime }\left( z\right) +\left[ A+\frac{B}{z}+\frac{C}{z-1}\right]
H^{\prime }\left( z\right) +\frac{ADz-h}{z\left( z-1\right) }H\left(
z\right) =0,  \label{35}
\end{equation}

The

After matching Eqs. (33, 34) with Eq. (35), one can get the following
correspondences.

a) For Eq. (33),

\begin{equation}
A=-4a^{2}m,\text{ \ \ }B=\frac{1}{2}+m+2a^{2}m,\text{ \ \ }C=\frac{1}{2}%
-(m+2a^{2}m),\text{ \ \ }D=0\text{ and }h=\lambda ^{2}.  \label{36}
\end{equation}

b) For Eq. (34),

\begin{equation}
A=4a^{2}m,\text{ \ \ }B=\frac{1}{2}-(m+2a^{2}m),\text{ \ \ }C=\frac{1}{2}%
+m+2a^{2}m,\text{ \ \ }D=0\text{ and }h=\lambda ^{2}.  \label{37}
\end{equation}

Determining when the solutions of a confluent Heun equation are expressible
in terms of more familiar functions would obviously be useful. Expansion of
solutions to the confluent Heun equation in terms of hypergeometric and
confluent hypergeometric functions are studied in detail by \cite{R12}. In 
\cite{R12}, it is also shown that the confluent Heun's functions can be
normalized to constitute a group of orthogonal complete functions. Here, as
an example, we follow the intermediate steps in \cite{R12} (page 102) in
order to express the solutions of Eq. (35) with $D=0$ in terms of the
hypergeometric functions. The transformation between the confluent Heun
function and the hypergeometric function is given with the Floquet expansion 
\cite{R12,R14}. Namely,

\begin{equation}
H_{j}(z)=\overset{\infty }{\underset{n=-\infty }{\dsum }}g_{n}\text{ }%
_{2}F_{1}(\xi _{1},\xi _{2};B;z),  \label{38}
\end{equation}

where

\begin{equation}
\xi _{1}=-n-\nu _{j}\text{ and }\xi _{2}=n+\nu _{j}+C+B-1,  \label{39}
\end{equation}

and $\nu _{j}$\ are known as the Floquet exponents. The coefficients $g_{n}$
satisfy a three-term recurrence relation

\begin{equation}
\Lambda _{n}g_{n-1}+Q_{n}g_{n}+\Phi _{n}g_{n+1}=0,  \label{40}
\end{equation}

where

\begin{equation*}
\Lambda _{n}=Ab_{n-1,n},
\end{equation*}

\begin{equation*}
Q_{n}=\xi _{1}\xi _{2}-h+\frac{A}{2}b_{n,n},
\end{equation*}

\begin{equation}
\Phi _{n}=Ab_{n+1,n},  \label{41}
\end{equation}

and the coefficients, $b_{n-1,n}$, $b_{n,n}$ and $b_{n+1,n}$ expressed in
terms of the parameters $\xi _{1},\xi _{2},$ are given explicitly by \cite%
{R12} .(The only difference between our notation and the Ronveaux's notation
is $B\equiv \gamma $ ).

Finally, it should be noted that one can see a brief analysis of the
confluent Heun equation as well as with its power series solution and
polynomial solution in \cite{R18}.

\section{Reduction of the angular equation into a set of linear first order
differential equations: The case with mass}

In this section, we shall reduce the angular Eqs. (16, 17) into a set of
linear set of first order differential equations for the case of Dirac
particle with mass. To this end, let us make the following substitutions
into Eqs. (16, 17),

\begin{equation}
A_{1}\left( \theta \right) =\left[ A_{0}\left( \theta \right) +iB_{0}\left(
\theta \right) \right] e^{-\int \left( \frac{\cot \theta }{2}-\frac{%
a^{2}\sin (2\theta )}{4F}\right) d\theta },  \label{42}
\end{equation}%
\begin{equation}
A_{3}\left( \theta \right) =\left[ M_{0}\left( \theta \right) +iN_{0}\left(
\theta \right) \right] e^{-\int \left( \frac{\cot \theta }{2}-\frac{%
a^{2}\sin (2\theta )}{4F}\right) d\theta },  \label{43}
\end{equation}

where $A_{0}\left( \theta \right) ,B_{0}\left( \theta \right) ,M_{0}\left(
\theta \right) $ and $N_{0}\left( \theta \right) $ are real functions. After
separating the real and the imaginary parts, one obtains a set of first
order differential equations 
\begin{equation*}
A_{0}^{\prime }\left( \theta \right) -\frac{mF}{\sin \theta }A_{0}\left(
\theta \right) +\frac{q\sin \theta }{4F}B_{0}\left( \theta \right) =\lambda
M_{0}\left( \theta \right) -\mu ^{\ast }\sqrt{F}N_{0}\left( \theta \right) ,
\end{equation*}%
\begin{equation*}
B_{0}^{\prime }\left( \theta \right) -\frac{mF}{\sin \theta }B_{0}\left(
\theta \right) -\frac{q\sin \theta }{4F}A_{0}\left( \theta \right) =\lambda
N_{0}\left( \theta \right) -\mu ^{\ast }\sqrt{F}M_{0}\left( \theta \right) ,
\end{equation*}%
\begin{equation*}
M_{0}^{\prime }\left( \theta \right) +\frac{mF}{\sin \theta }M_{0}\left(
\theta \right) +\frac{q\sin \theta }{4F}N_{0}\left( \theta \right) =-\lambda
A_{0}\left( \theta \right) -\mu ^{\ast }\sqrt{F}B_{0}\left( \theta \right) ,
\end{equation*}%
\begin{equation}
N_{0}^{\prime }\left( \theta \right) +\frac{mF}{\sin \theta }N_{0}\left(
\theta \right) -\frac{q\sin \theta }{4F}M_{0}\left( \theta \right) =-\lambda
B_{0}\left( \theta \right) -\mu ^{\ast }\sqrt{F}A_{0}\left( \theta \right) .
\label{44}
\end{equation}

Introducing a new variable $x=\cos \theta $, one can remove the
trigonometric functions in the set of first order differential equations
(45). But solving the entire system analytically does not seem possible. To
our knowledge, in the literature such a system does not exist. Nevertheless,
one can analyze the system via an appropriate numerical technique, which may
need an advanced computational study.

\section{Conclusion}

In this paper, our target was not only to separate the Dirac equation for a
test spin-$\frac{1}{2}$ particle in the rotating electromagnetic spacetime
(RBR), but to explore exact solutions as well. By this way, we wanted to
make a contribution to the wave mechanical aspects of the Dirac particles in
the RBR geometry.

Due to the metric functions, which are only functions of the angular
variable $\theta $, the angular part of the Dirac equation in the RBR
background is the harder part to be tackled compared with the axial part.
Another advantage of the axial part is that the axial equations do not
involve the mass term. These simplifications in the axial equations guided
us to obtain the general solution of the axial part in terms of the
hypergeometric functions. On the other hand, although we could not obtain
the general analytic solution of the angular part, we succeeded to overcome
the difficulties in the angular part in the massless \ case and obtained an
exact solution of it in terms of the Heun polynomials. Inclusion of mass
prevents us to obtain an analytic solution for the angular part. As an
alternative way to the analytic solution, in the last section we showed that
the angular part could be written as a set of first order differential
equations, which are suitable for numerical investigations.

Finally, the study of the charged Dirac particles in the RBR spacetime may
reveal more information compared to the present case. This is going to be
our next problem in the near future.
\begin{verbatim}
Acknowledgement
\end{verbatim}

We would like to thank Prof. M. Halilsoy for his helpful comments during our
work.

\end{document}